# On the sensitivity of the diffusion MRI signal to brain activity in response to a motor cortex paradigm

A. De Luca[1], L. Schlaffke[2], J.C.W. Siero[3,4], M. Froeling[3], A. Leemans[1]

[1]Image Sciences Institute, UMC Utrecht, Utrecht, The Netherlands

[2]Department of Neurology, BG-University Hospital Bergmannsheil, Ruhr-University Bochum, Bochum, Germany

[3]Department of Radiology, UMC Utrecht, Utrecht, The Netherlands

[4]Spinoza Centre for Neuroimaging, Amsterdam, The Netherlands

Corresponding Author:

Alberto De Luca

Heidelberglaan 100, 3584CX, Utrecht, The Netherlands

alberto@isi.uu.nl

Highlights:

- We propose an acquisition scheme for disentangling multiple signal contributions in diffusion fMRI
- The measured signal exhibits sensitivity to brain activity in the motor cortex;
- Simultaneous signal changes take place in the hindered and perfusion compartments;
- The sensitivity of the diffusion MRI signal to task is similar to that of gradient echo fMRI;

Abbreviations: MRI: magnetic resonance imaging; dMRI: diffusion MRI; dfMRI: diffusion functional MRI; ADC: apparent diffusion coefficient; ADC-fMRI: ADC based functional MRI; IVIM: intra-voxel incoherent motion; IVIM-fMRI: IVIM based functional MRI; GE-BOLD: gradient-echo blood oxygenation level dependent; AC: activation core;

Conflicts of interest: none

## Abstract

Diffusion functional MRI (dfMRI) is a promising technique to map functional activations by acquiring diffusion-weighed spin-echo images. In previous studies, dfMRI showed higher spatial accuracy at activation mapping compared to classic functional MRI approaches. However, it remains unclear whether dfMRI measures result from changes in the intra-/extracellular environment, perfusion and/or $T_2$ values. We designed an acquisition/quantification scheme to disentangle such effects in the motor cortex during a finger tapping paradigm. dfMRI was acquired at specific diffusion weightings to selectively suppress perfusion and free-water diffusion, then times series of the apparent diffusion coefficient (ADC-fMRI) and of the perfusion signal fraction (IVIM-fMRI) were derived. ADC-fMRI provided ADC estimates sensitive to changes in perfusion and free-water volume, but not to $T_2/T_2^*$ values. With IVIM-fMRI we isolated the perfusion contribution to ADC, while suppressing $T_2$ effects. Compared to conventional gradient-echo BOLD fMRI, activation maps obtained with dfMRI and ADC-fMRI had smaller clusters, and the spatial overlap between the three techniques was below 50%. Increases of perfusion fractions were observed during task in both dfMRI and ADC-fMRI activations. Perfusion effects were more prominent with ADC-fMRI than with dfMRI but were significant in less than 25% of activation ROIs. Taken together, our results suggest that the sensitivity to task of dfMRI derives from a decrease of hindered diffusion and an increase of the pseudo-diffusion signal fraction, leading to different, more confined spatial activation patterns compared to classic functional MRI.

# Introduction

Diffusion MRI (dMRI) is a non-invasive technique sensitive to the in vivo displacement of water molecules. Despite the dynamic nature of the signal, dMRI is mostly regarded as a structural technique providing static and reproducible snapshots of the imaged tissue [Acheson et al., 2017; Grech-Sollars et al., 2015]. The dMRI signal carries information about different components of tissue microstructure. At strong diffusion weightings, the signal is mainly informative of the intra and extra-cellular environments [Winston, 2012], whereas at low diffusion weightings, it is sensitive to intra-voxel incoherent motion (IVIM) phenomena, including perfusion [Le Bihan et al., 1988] and free water diffusion [Pasternak et al., 2009; Pierpaoli and Jones, 2004]. In the brain, the IVIM model has been used to quantify perfusion changes during $CO_2$ challenges [Federau et al., 2012], or in presence of disease [Iima and Le Bihan, 2016] as cancer [Keil et al., 2017]. The concept of IVIM and its applications suggest sensitivity of the dMRI signal acquired at low (i.e., $b \leq 200$ s/mm$^2$) to moderate b-values (i.e., $b \leq 500$ s/mm$^2$) to physiological dynamics in the vascular and free-water pools. Such sensitivity applies also to the non-diffusion weighted images, which are the reference images for most diffusion experiments.

At intermediate diffusion weightings (i.e., $1400 \geq b \geq 1000$ s/mm$^2$), the dMRI signal measured in the brain originates mainly from hindered (intra- and extracellular) components, where structure and not function is assumed to be fairly constant over short time spans. In other words, if dMRI is assumed to be insensitive to function, it is also reasonable to consider that it is static, implying that the signal should not change over time beyond experimental noise. Nevertheless, an increasing number of reports is challenging the static nature of dMRI at strong diffusion weightings in the brain. Darquié and colleagues [Darquié et al., 2001] originally observed that visual stimuli administered during a dMRI experiment caused small but reproducible alterations of the apparent diffusion coefficient (ADC). In their experiment, the authors repeatedly acquired data at both low ($b = 200$ s/mm$^2$) and intermediate diffusion weightings ($b = 1400$ s/mm$^2$). In a later experiment, which included more diffusion weightings [Le Bihan et al., 2006], the authors proposed a bi-exponential formulation based on a slow and fast pool model, and observed a 1.7% expansion of the slower diffusion pool during visual stimuli. In the conclusions of their work, cell swelling after neuronal firing was proposed as an explanation of the findings. Other reports have independently confirmed the potential of diffusion functional MRI (dfMRI) in terms of improved spatial localization of brain activations as compared to standard gradient-echo blood oxygenation level-dependent fMRI (GE-BOLD) [Song et al., 2002], and showed that it well represents underlying neuronal activity in rats [Nunes et al., 2019]. Furthermore, the response function (RF) of the dfMRI signal, a mathematical description that relates the stimulation paradigm to signal changes, has been shown to have a shorter time to peak compared to GE-BOLD [Aso et al., 2009], which can be considered to be supporting the neuronal firing theory.

Despite these observations, most of the mechanisms of dfMRI, as well as its feasibility and usefulness, are yet to be investigated. Miller et al. [Miller et al., 2007] challenged the neuronal firing theory, suggesting a major role of the "$T_2$ shine through" effect [Provenzale et al., 1999] on the dfMRI signal. Furthermore, previous studies employed either low [Song et al., 2002] or intermediate to strong diffusion weightings [Le Bihan et al., 2006; Darquié et al., 2001; Nicolas et al., 2017; Williams et al., 2014], which are sensitive to different features of the microstructural environment, and evidence beyond the visual cortex is limited [Aso et al., 2013; Song et al., 2002].

In this work, we investigated whether the dfMRI signal measured at increasing diffusion weightings, and the derived apparent diffusion coefficient (ADC) fMRI (ADC-fMRI), and intra-voxel incoherent motion fMRI (IVIM-fMRI) signals, are sensitive to brain activations in response to a finger tapping paradigm. In such case, we hypothesize that it is possible to disentangle the contributions of hindered diffusion, free-water diffusion and blood perfusion to the observed signal changes, by taking advantage of multiple diffusion weightings acquired in our dfMRI experiments.

# Methods

In this study we performed fMRI acquisitions while eliciting brain activations in the motor cortex with a finger-tapping task [Akhlaghi et al., 2012]. Being well studied, straightforward to implement, and consisting of only two conditions (rest vs task), this paradigm represented an excellent starting point for the investigation of the dfMRI signal in the motor cortex.

## Dataset

Seven subjects (24 ± 3 years, 4 females) underwent a 3T MRI session. The experiment was approved by the local ethical committee, and informed written consent was obtained from all subjects. The data that support the findings of this study are available on request from the corresponding author. The data are not publicly available due to privacy or ethical restrictions.

## MRI acquisition

Data were acquired on a research dedicated Philips Ingenia CX scanner (Philips Medical System NV) with multi-band (MB) imaging capabilities [Setsompop et al., 2012] and a 32ch head coil. The acquisition protocol consisted of anatomical 3D $T_1$-weighted imaging, a GE-BOLD, and two dfMRI acquisitions. The main imaging parameters of each sequence are reported in Table 1. The dfMRI sequence was spin-echo EPI with monopolar Stejskal-Tanner gradients of varying amplitude. The characteristic times of the gradients were $\Delta/\delta = 48.7/23.7\text{ms}$ for the first run, and $\Delta/\delta = 52.8/27.6\text{ms}$ for the second run. The echo-time of two dfMRI sequence was set to shortest, resulting in a 6ms difference to accommodate the longer diffusion weightings employed in the second run. The second dfMRI run of one subject was not completed due to time constraints.

| Sequence | Resolution (mm³) | MB | SENSE | TE (ms) | TR (ms) | Slices | FOV (mm³) | b-values (s/mm²) | Bandwidth in PE (Hz) | Duration |
|---|---|---|---|---|---|---|---|---|---|---|
| $T_1$-w | 1x1x1 | - | 2,2.6 | 3.7 | 8 | 180 | 256x256x180 | - | 191.5 | 3m54s |
| GE-BOLD | 2.5x2.5x3 | 2 | 2 | 30 | 2000 | 52 | 240x240x156 | - | 34.2 | 6m40s |
| dfMRI 1 | 2.5x2.5x4 | 2 | 2 | 85 | 8000 | 14 | 240x240x56 | 0,300,800 | 35.0 | 6m50s |
| dfMRI 2 | 2.5x2.5x4 | 2 | 2 | 91 | 8000 | 14 | 240x240x56 | 0,300,1200 | 33.9 | 6m50s |

Table 1 – Imaging parameters of the sequences employed in this study. No slice gap was employed for any of the sequences. All functional acquisitions were performed with echo planar imaging readout. Data at b = 300, 800, 1200 s/mm$^2$ were acquired with gradients along three orthogonal directions aligned with the scanner axes. MB: multi-band; SENSE: sensitivity encoding parallel imaging acceleration; TE: echo time; TR: repetition time; FOV: field of view; PE: phase encoding.

## Functional acquisitions

To elicit motor activation, we implemented a previously reported experiment design [Nicolas et al., 2017], which alternates six repetitions of rest and activation blocks with 32 seconds duration (384 s in total). Instructions were presented on a video device installed in the MR scanner room, and the start condition (task or rest) was pseudo-randomized across subjects (5 subjects started with task blocks, whereas 2 with rest blocks). Three functional datasets were acquired for each subject. The first acquisition was a GE-BOLD sequence featuring 16 dynamics repeated over six task and six rest blocks, for a total of 192 volumes. The second and the third acquisitions were dfMRI sequences with four dynamics per diffusion weighting, repeated in six task and six rest blocks, for a total of 48 volumes per diffusion weighting and 48 non-weighted images (b = 0 s/mm$^2$), which are also referred to as Spin-Echo BOLD (SE-BOLD) [Glielmi et al., 2010]. The acquisition of multiple diffusion weightings resulted in less dynamics per block compared to a GE-BOLD acquisition. However, with these data we can investigate different microstructural components, as explained in the following section.

## Separating physiological contributions with dfMRI, ADC-fMRI and IVIM-fMRI

The GE-BOLD signal is $T_2^*$ weighted and is sensitive to changes in both blood volume, blood flow and oxygenation. The spin-echo dfMRI images are sensitive to changes in $T_2$ and in (blood) water compartment volumes (the vasculature, intra- and extra-cellular water, and free water), according to the multi-compartment signal model we will use here. The signal $S$ measured at time $t_j$, echo time $T_E$, and diffusion weighting b, can be expressed as:

$$S(t_j) = S_0 \sum_i f_i e^{-\frac{T_E}{T_{2,i}(t_j)}} e^{-bD_i(t_j)},$$

where $S_0$ is the non-weighted signal, whereas $f_i$ is the signal fraction, $T_{2,i}$ the $T_2$ value, and $D_i$ the diffusion coefficient of the i-th component. Adjusting the diffusion weighting in the dfMRI experiment allows one to selectively suppress specific contributions. Data acquired at b = 0 s/mm$^2$ (SE-BOLD) are sensitive to all tissue components, whereas data at b = 300 s/mm$^2$ are considered largely free of contributions from large vessels (e.g. signal change < 5% for spins diffusing faster than 10x10$^{-3}$mm$^2$/s) [Federau et al., 2015; Finkenstaedt et al., 2017; H. et al., 2011]. At stronger diffusion weightings, as b = 800 s/mm$^2$ and b = 1200 s/mm$^2$, the signal is considered to be insensitive to most vascular contributions (e.g. signal change < 5% for spins diffusing faster than 3.7x10$^{-3}$ mm$^2$/s and 2.5x10$^{-3}$ mm$^2$/s, respectively). Data acquired at b = 1200 s /mm$^2$ are also negligibly sensitive to contributions from free water diffusion, as the dMRI signal is attenuated by over 96% at that diffusion weighting.

If the tissue composition is mono-compartmental, i.e. i = 1 in the above equation, all dfMRI data are sensitive to $T_2$ changes independently from the applied diffusion weighting. However, if $T_2$ changes take place only in specific components, for instance only in the vascular network, dfMRI data acquired at b $\geq$ 300 s/mm$^2$ should be insensitive to such effects. In addition to the $T_2$ sensitivity, the EPI-readout of the used dfMRI sequence is to some extent sensitive to $T_2^*$ changes – and hence to the BOLD contrast – but to a much lesser extent than conventional $T_2^*$ weighted GE-BOLD [Goense and Logothetis, 2006].

If the acquired dfMRI data contains at least two diffusion weightings, it is possible to compute the apparent diffusion coefficient (ADC) over time (ADC-fMRI). ADC-fMRI is regarded as insensitive to $T_2$ and $T_2^*$ effects, as such dependencies cancel out in the ADC equation. However, this is only true for a mono-component tissue formulation (i=1). If multiple components coexist in one voxel, ADC-fMRI becomes sensitive to factors affecting the b = 0 s/mm$^2$ image, such as perfusion and free water volume changes, among others.

If the dfMRI experiment features three or more diffusion weightings with appropriate values, it is possible to subdivide the signal contributions into vascular and non-vascular components, applying the IVIM-fMRI model in a bi-exponential formulation (i = 2):

$$\frac{S}{S_0} = f_{IVIM} e^{-bD^*} + (1 - f_{IVIM}) e^{-bMD_{IVIM}}.$$

In the above equation, $f_{IVIM}$ represents the signal fraction of a pool including both pseudo-diffusion and free-water contributions with diffusion coefficient $D^*$, whereas $MD_{IVIM}$ reflects the contributions of both intra- and extra-cellular diffusion changes. Under such assumptions, the temporal series of $f_{IVIM}$ (in the following, referred to as $f_{IVIM}$-fMRI) reflects the volumes changes related to free-water/perfusion and is theoretically less sensitive to $T_2$ and $T_2^*$ contamination than ADC-fMRI, whereas the temporal series of $MD_{IVIM}$ (in the following, referred to as $MD_{IVIM}$-fMRI) summarizes hindered diffusion changes.

## Data pre-processing

GE-BOLD data of each subject was corrected for subject motion using the FSL [Jenkinson et al., 2012] tool MCFLIRT [Jenkinson et al., 2002], realigning all volumes to the first volume. Individual GE-BOLD data were used as the space of reference for all functional analyses.

The processing pipeline applied to dfMRI data is elucidated in Figure 1. The two dfMRI series were concatenated, corrected for subject motion and eddy currents using ExploreDTI [Leemans et al., 2009], aligning all dfMRI volumes to the first b = 0 s/mm$^2$ image. The first non-weighted image was registered to the individual fMRI space using a non-linear b-spline transformation restricted to the phase-encoding direction [Klein et al., 2010], then all the data were transformed with a single interpolation step. Brain extraction was performed on the dfMRI data with FSL BET [Smith, 2002]. The dMRI data corresponding to each diffusion weighting (b-value) were geometrically averaged. To determine the SNR of the data, a homogeneous region was manually delineated on an axial slice of each subject. The SNR was determined as ratio between spatial average and standard deviation and corrected for the Rician nature of the noise [Gudbjartsson and Patz, 1995]. The temporal SNR (tSNR) of all dataset was determined within the same ROI as temporal average divided by temporal standard deviation.

The FreeSurfer reconstruction pipeline [Dale et al., 1999] was applied to $T_1$-weighted data to derive the GM/WM interface used for the graphical representations. Further, the FSL pipeline "fsl_anat" [Jenkinson et al., 2012; Smith, 2002; Zhang et al., 2001] was applied to each T1 image to derive segmentation of GM and WM, which were registered to the fMRI space with Elastix [Klein et al., 2010] using a b-spline registration [Klein et al., 2007].

## dfMRI, ADC-fMRI, IVIM-fMRI processing

We performed Z-normalization of the dfMRI data, i.e., we subtracted from each time series its average value and divided it by the standard deviation, independently for each diffusion weighting. We then concatenated the normalized data to maximize the statistical power in the subsequent analyses.

The ADC-fMRI estimates were computed by combining the data at b = 0 s/mm$^2$ with the data at b = 300 s/mm$^2$ (ADC-fMRI$^{300}$), b = 800 s/mm$^2$ (ADC-fMRI$^{800}$), and b = 1200 s/mm$^2$ (ADC-fMRI$^{1200}$) separately, using the ordinary linear least-squares fit (MATLAB R2016b, The Mathworks Inc). Each ADC-fMRI series was normalized independently and then concatenated.

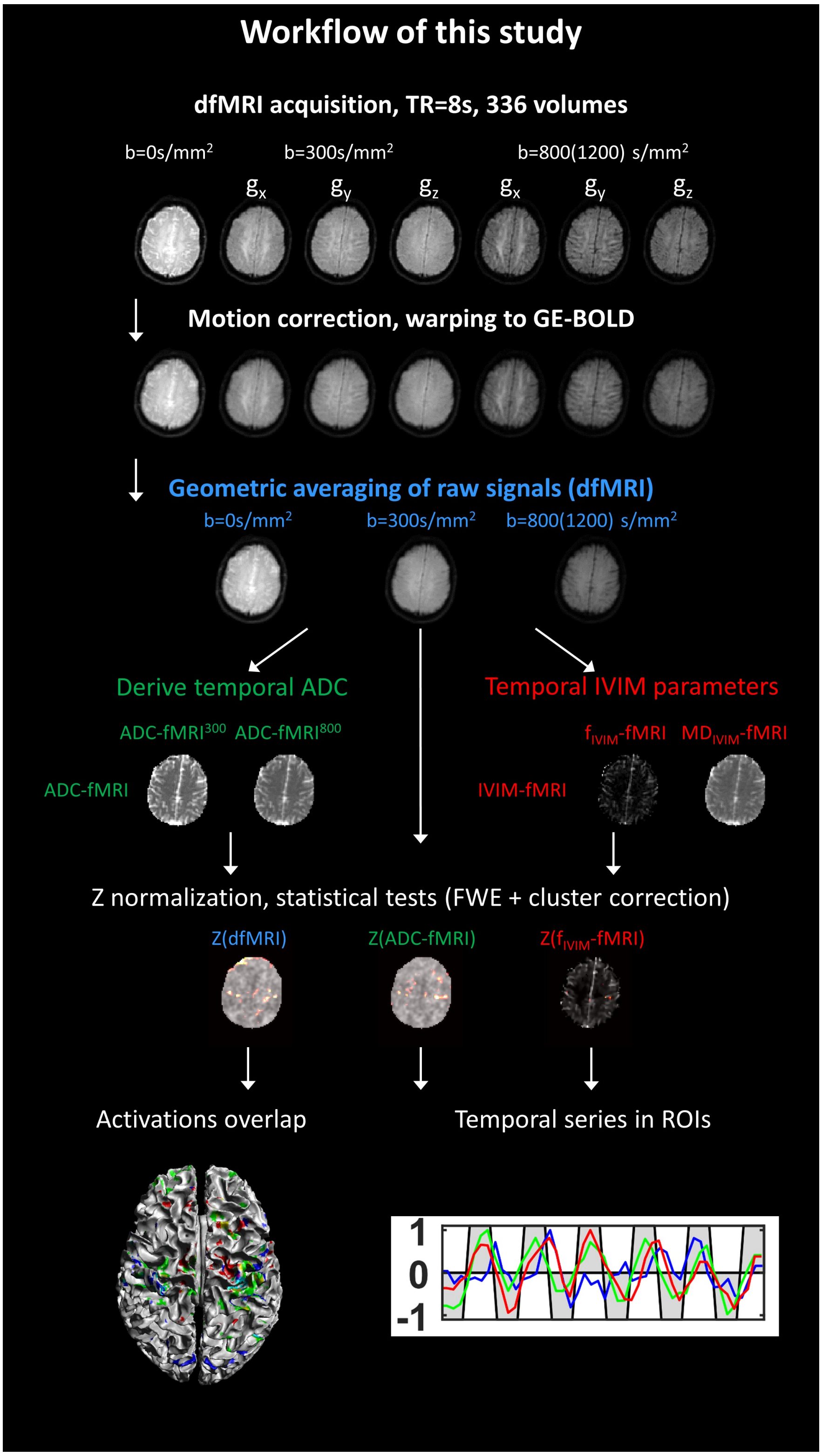

Figure 1: Workflow of this study. dfMRI data are firstly processed to attenuate motion and eddy currents related artefacts, then warped to the individual GE-BOLD space, which is used as standard space for all analysis. Data are geometrically averaged per diffusion weighting and used 1) directly for activation mapping or 2) to derive ADC-fMRI and IVIM-fMRI. After activations are individually mapped, temporal series of the signals in the activation ROIs are computed. Acronyms: $g_{x,y,z}$: diffusion gradient along the x, y or z axis; FWE: family-wise error; $TTP_{1,2}$: time-to-peak of the two Gamma functions.

The IVIM-fMRI fit of the data was performed by using a bi-exponential formulation [Cho et al., 2015], and a segmented fit method [Sigmund et al., 2012]. Accordingly, perfusion corrected MD values were computed using data at two diffusion weightings, while the normalized difference between the measured and the estimated non-weighted signal represented the perfusion fractions. Values of perfusion fractions ($f_{IVIM}$-fMRI) and of the perfusion corrected mean diffusivity ($MD_{IVIM}$-fMRI) over time were computed for both the first ($f_{IVIM}$-fMRI$^{300,800}$, $MD_{IVIM}$-fMRI$^{300,800}$) and the second acquisition ($f_{IVIM}$-fMRI$^{300,1200}$, $MD_{IVIM}$-fMRI$^{300,1200}$). The abovementioned approaches are schematically summarized in Figure 1.

The merged Z(dfMRI), Z(ADC-fMRI) and Z($f_{IVIM}$-fMRI$^{300,800}$) time-series featured 288, 192, and 48 volumes, respectively.

## Activation ROIs mapping

Task induced activations were mapped on the GE-BOLD data using FSL FEAT with cluster correction for multiple comparisons. Voxel-wise Z-statistics were computed by convolving the paradigm design with the default hemodynamic response function (HRF), a gamma function with average delay 3s and standard deviation 6s [Glover, 1999]. The Z-statistics were corrected for multiple comparisons using a first level Z-threshold equal to 0.5 and a cluster p-value equal to 0.05, and finally thresholded at $Z \geq 2.3$ [Nicolas et al., 2017]. The statistical analysis of Z(dfMRI), Z(ADC-fMRI), Z($MD_{IVIM}$-fMRI) and Z($f_{IVIM}$-fMRI) was performed similarly to that of GE-BOLD but employing a box-car response function instead of the default HRF. Flattened projection maps of the Z-scores of the spatial activations with GE-BOLD, dfMRI, ADC-fMRI can be found in the Supporting Information.

### Time-series in activation ROIs

Statistically significant ROIs of Z(dfMRI) and Z(ADC-fMRI) were thresholded above 70% of their peak value to determine their activation core (AC) [Nicolas et al., 2017], here defined as AC(dfMRI) and AC(ADC-fMRI), respectively. dfMRI at each diffusion weighting, ADC-fMRI and GE-BOLD were corrected for linear trends, divided by their maximum value, spatially averaged within the AC ROIs and temporally processed with a three-points moving average filter, to mitigate high frequency noise.

Further, the Z-scores of the average GE-BOLD, dfMRI and ADC-fMRI signals of each subject were computed and overlaid to investigate the sensitivity of the signals to task, as well as to qualitatively observe temporal aspects – e.g. lags in their response. To compute the Z-scores, the signals were corrected for linear trends and divided by their standard deviation after demeaning.

Relative percent changes of dfMRI, ADC-fMRI and $f_{IVIM}$-fMRI between the task and rest conditions were derived for each subject. The signals were averaged within the ROIs and over time within the corresponding blocks, then statistical significance was assessed with Z-tests.

### Spatial overlap of activation ROIs

The spatial agreement between two spatial activation maps, referred as A and B, was quantified using the overlap metric, which is defined as:

$$\mathrm{OVERLAP}(A, B) = \frac{|A \cap B|}{\min(|A|, |B|)},$$

with the operator |A| indicating the volume of A.

Comparisons were performed between GE-BOLD, dfMRI and ADC-fMRI. Furthermore, the effect of perfusion on dfMRI and ADC-fMRI was quantified by computing their overlap with $f_{IVIM}$-fMRI activations. Projection maps of the Z-scores of the overlaps between GE-BOLD, dfMRI and ADC-fMRI were computed by summing their values along the feet-head direction, to qualitatively appreciate their extents in a planar format. These maps can be found in the Supporting Information. Finally, to understand whether dfMRI and ADC-fMRI activations were specific of GM or WM processes, we computed the overlap between the segmentations derived from the T1 images and GE-BOLD, dfMRI and ADC-fMRI activations.

## Results

### dfMRI

The average SNR and tSNR of the acquired dataset are reported in Table 2. Both the SNR and the tSNR of dfMRI data were sufficient and comparable to GE-BOLD up to b = 800s/mm$^2$, while ADC-fMRI provided remarkably lower values. Figure 2 shows the 3D mapping on the WM/GM interface of the activations detected in correspondence of the task > rest condition with dfMRI, GE-BOLD and their overlap. Bilateral activation in the primary motor cortex area is observed on the dfMRI activation maps of all subjects, whereas the supplementary motor area is detected only on 2 out of 7 subjects. The activation maps with dfMRI include some artefactual patterns and have smaller activation clusters than those with GE-BOLD. The overlap between statistically significant voxels of dfMRI and GE-BOLD is 45 $\pm$ 14%. Axial projections of the activation maps and of the spatial overlap are shown in Supporting Figure S1. dfMRI activations overlapped more with GM than WM, 41$\pm$7% vs 31$\pm$10%, respectively.

| Signal | AC(dfMRI) | | | AC(ADC-fMRI) | | |
|---|---|---|---|---|---|---|
| | Average value | Signal change | p-value | Average value | Signal change | p-value |
| GE-BOLD | - | +1.02 ± 0.64 | **< 0.01** | - | +0.72 ± 0.53 | **0.01** |
| dfMRI-0 | - | +0.51 ± 0.20 | **< 0.01** | - | +0.47 ± 0.24 | **< 0.01** |
| dfMRI-300 | - | +0.34 ± 0.23 | **< 0.01** | - | -0.06 ± 0.18 | 0.38 |
| dfMRI-800 | - | +0.44 ± 0.24 | **< 0.01** | - | -0.13 ± 0.31 | 0.31 |
| dfMRI-1200 | - | +0.53 ± 0.19 | **< 0.01** | - | -0.22 ± 0.18 | **0.03** |
| ADC-fMRI-300 | $(1.11 \pm 0.08)x10^{-3}mm^2/s$ | +0.55 ± 0.26 | **< 0.01** | $(1.10 \pm 0.11)x10^{-3}mm^2/s$ | +2.17 ± 1.15 | **< 0.01** |
| ADC-fMRI-800 | $(0.92 \pm 0.08)x10^{-3}mm^2/s$ | +0.06 ± 0.16 | 0.37 | $(0.93 \pm 0.09)x10^{-3}mm^2/s$ | +1.14 ± 0.57 | **< 0.01** |
| ADC-fMRI-1200 | $(0.85 \pm 0.07)x10^{-3}mm^2/s$ | +0.05 ± 0.21 | 0.58 | $(0.83 \pm 0.06)x10^{-3}mm^2/s$ | +1.18 ± 0.76 | **0.01** |
| $MD_{IVIM}$-fMRI-300-800 | $(0.83 \pm 0.08)x10^{-3}mm^2/s$ | -0.28 ± 0.24 | **0.02** | $(0.84 \pm 0.08)x10^{-3}mm^2/s$ | +0.46 ± 0.58 | 0.08 |
| $MD_{IVIM}$-fMRI-300-1200 | $(0.78 \pm 0.06)x10^{-3}mm^2/s$ | -0.26 ± 0.25 | **0.05** | $(0.75 \pm 0.08)x10^{-3}mm^2/s$ | +0.56 ± 0.53 | **0.05** |
| $f_{IVIM}$-fMRI-300-800 | 0.07 ± 0.01 | +3.92 ± 1.81 | **< 0.01** | 0.07 ± 0.01 | +7.70 ± 3.67 | **< 0.01** |
| $f_{IVIM}$-fMRI-300-1200 | 0.09 ± 0.01 | +3.16 ± 2.07 | **0.01** | 0.09 ± 0.01 | +6.02 ± 2.53 | **< 0.01** |

Table 2 – Values at rest of the time-series of dfMRI, ADC-fMRI, $MD_{IVIM}$-fMRI $f_{IVIM}$-fMRI in AC(dfMRI) and AC(ADC-fMRI), and their signal change during task as compared to rest. The p-value refers to a two-sided t-test between the average values in the two conditions and is highlighted in bold when significant.

Time-series of GE-BOLD, dfMRI (for different diffusion weightings), $MD_{IVIM}$-fMRI and $f_{IVIM}$-fMRI averaged within AC(dfMRI) are reported in Figure 3. All diffusion weighted signals consistently show increases in correspondence of task execution followed by decreases at rest, similarly to GE-BOLD.

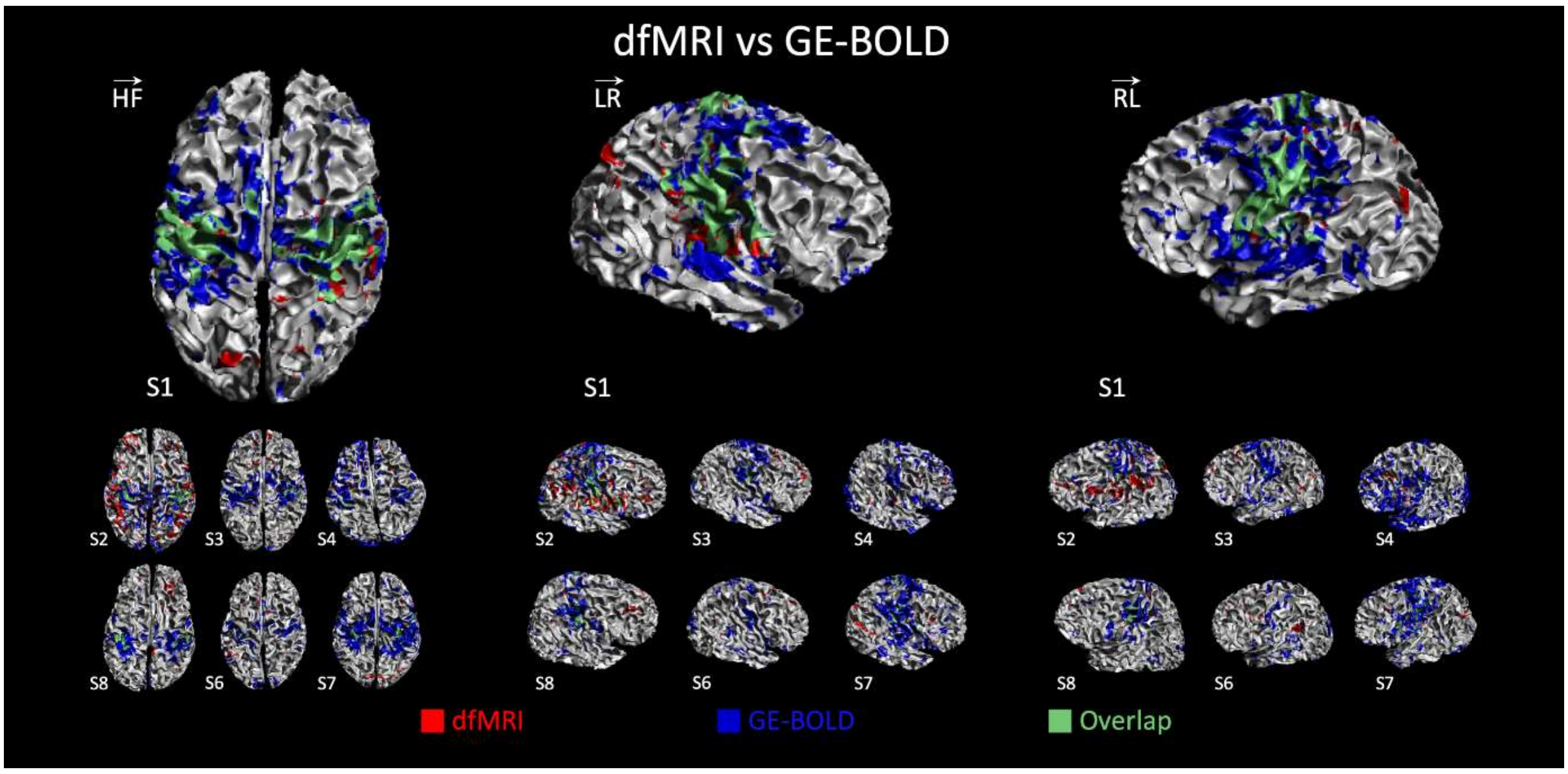

Figure 2: dfMRI vs GE-BOLD activation maps. Individual activation maps detected with dfMRI (red) as compared to GE-BOLD (blue), and their overlap (green) overlaid on the grey/white matter surface of each subject. Bilateral activation as response to finger tapping was observed on all subjects with both sequences. Activations with dfMRI were smaller than those with GE-BOLD. Spurious activations due to multi-band reconstruction artefacts can be spotted on the dfMRI activations of some subjects.

Although all signal changes are consistent with the task-rest paradigm, signals acquired at b = 800, 1200 s/mm$^2$ exhibit a larger number of artefacts than those at b = 300 s/mm$^2$. Considering the average signals within task and rest blocks, we observe significant increases of +0.51 $\pm$ 0.20% ($p=10^{-4}$), +0.34 $\pm$ 0.23% ($p = 10^{-3}$), +0.44 $\pm$ 0.24% ($p = 10^{-3}$) and +0.53 $\pm$ 0.19% ($p = 10^{-4}$) during the task for b = 0, 300, 800, and 1200 s/mm$^2$, respectively. Regarding ADC-fMRI signals in AC(dfMRI), only ADC-fMRI$^{300}$ showed significant signal increases during task compared to rest.

$MD_{IVIM}$-fMRI and $f_{IVIM}$-fMRI are very noisy at individual levels, but on average show response to task execution, with $f_{IVIM}$-fMRI$^{300\text{-}800}$ showing an increase between 4 and 5%, and $MD_{IVIM}$-fMRI a decrease during finger tapping. The average changes of the considered metrics from task to rest within AC(dfMRI), and their statistical significance, are reported in Table 2.

Activations detected with $f_{IVIM}$-fMRI appear noisier than those with dfMRI and characterized by smaller spatial extents. The overlap between activations from the two techniques is limited, with values equal to 15 $\pm$ 14%.

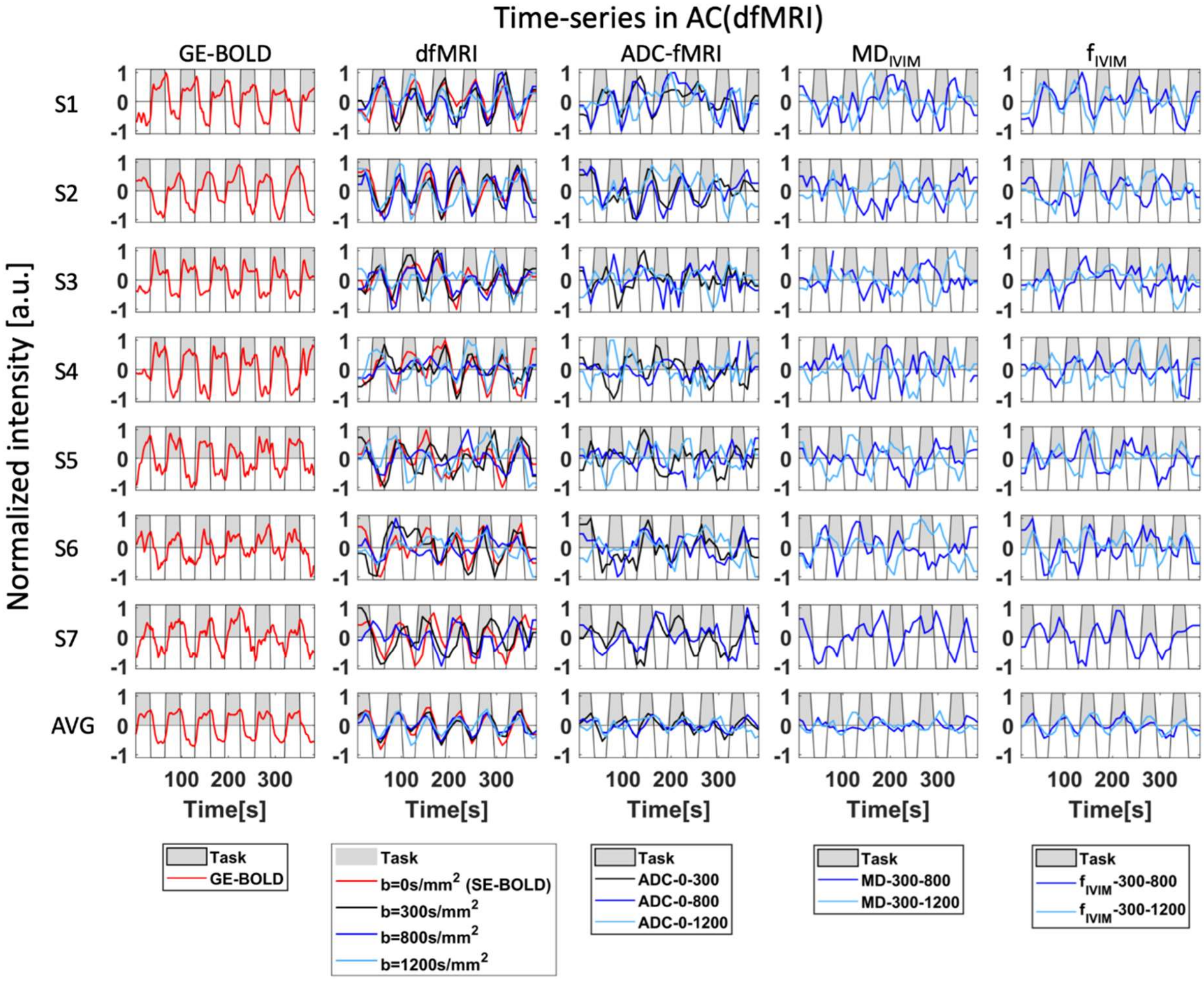

Figure 3: Time-series in AC(dfMRI). Normalized average time-series of GE-BOLD, dfMRI (for different diffusion weightings, red, black, blue and light blue solid lines), ADC-fMRI, $MD_{IVIM}$ and $f_{IVIM}$ of each subject (each row), and after group averaging (last row), within the dfMRI activation ROIs. Grey blocks correspond to task execution, whereas white blocks to rest. GE-BOLD and dfMRI showed increases and decreases in correspondence of task and rest, respectively. $f_{IVIM}$ showed synchronization with the task execution, but to a less extent than dfMRI. The $MD_{IVIM}$ signal was characterized by strong pseudo-random fluctuations, but its average variation suggested its decrease during task execution.

Figure 4 shows the Z-values of the time-series of GE-BOLD and dfMRI at multiple diffusion weightings. The sensitivity of the dfMRI signal to task is similar to that of GE-BOLD, with increases and decreases reaching $\pm$ 2 standard deviations of the signal variation. Considering the average inter-subject signals, changes in dfMRI at b = 1200 s/mm$^2$ slightly preceded those of SE-BOLD (b = 0 s/mm$^2$).

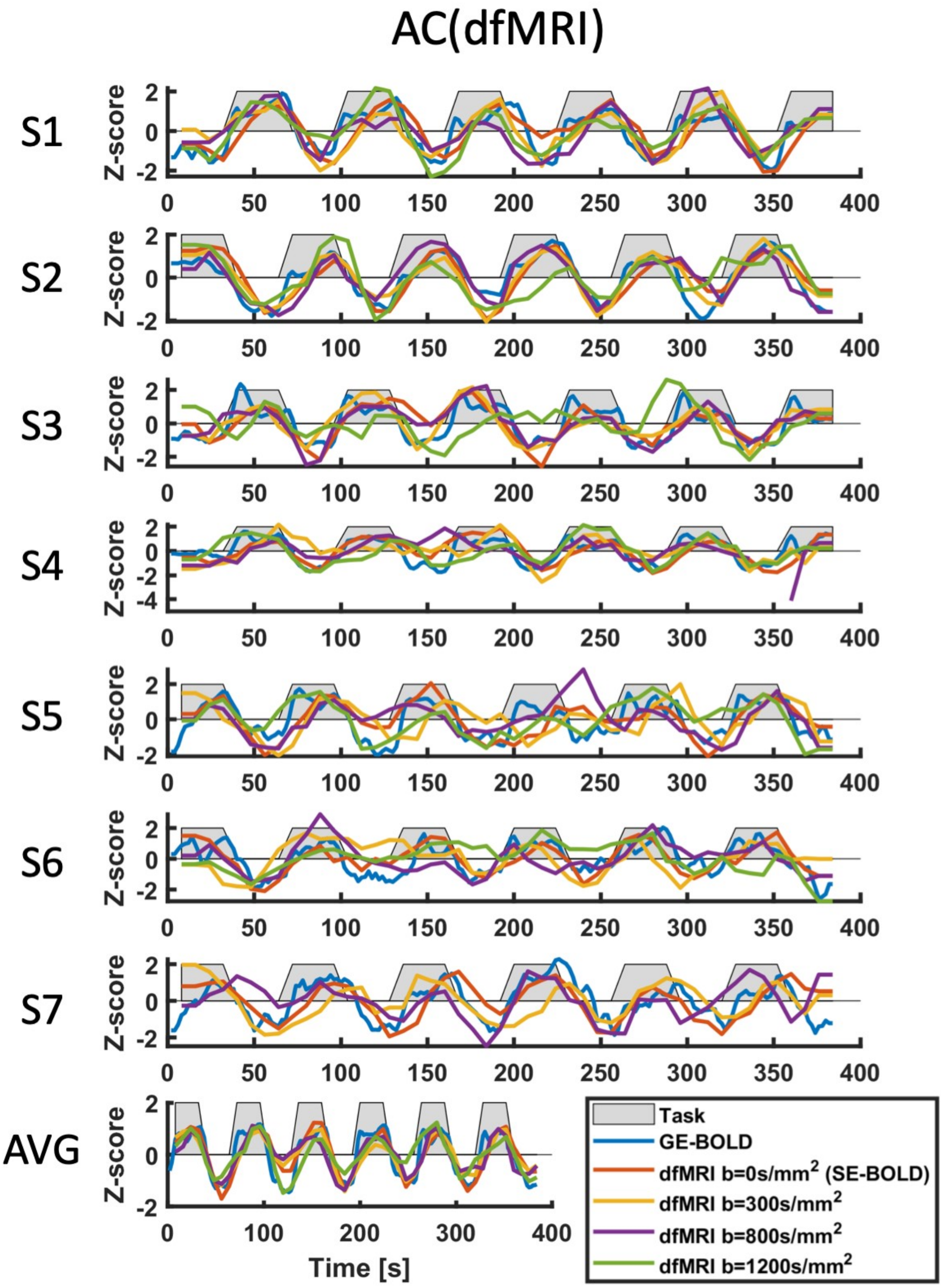

Figure 4: Sensitivity of GE-BOLD and dfMRI in AC(dfMRI). Z-score of the time-series of GE-BOLD and dfMRI (for different diffusion weightings, red, orange, purple solid lines) within the dfMRI activation core. The individual dfMRI signals showed signal changes up to 2 standard deviations in correspondence of task execution in line with GE-BOLD, irrespectively of the applied diffusion-weighting. The time-series suggest that dfMRI signals at b = 1200s/mm$^2$ exhibited slightly faster reactivity to task than SE-BOLD (b = 0 s/mm$^2$), whereas differences with GE-BOLD were minimal.

## ADC-fMRI

Figure 5 shows a 3D rendering of the functional activation maps detected with ADC-fMRI, GE-BOLD and their overlap on the GM/WM interface. Bilateral activations in the primary motor cortex are observed on all subjects with ADC-fMRI, but they have a smaller extension than those observed with dfMRI. Furthermore, activations with ADC-fMRI appear noisier compared to dfMRI. The overlap between Z(ADC-fMRI) and GE-BOLD is 33 ± 15%, while the overlap between Z(ADC-fMRI) and Z(dfMRI) is the lowest, 16 ± 10%. Axial projections of the activation maps of ADC-fMRI and of its spatial overlap with dfMRI and GE-BOLD are shown in Supporting Figure S1. ADC-fMRI activations were equally located in GM and WM, with overlap values of 38±6% and 38±4%, respectively.

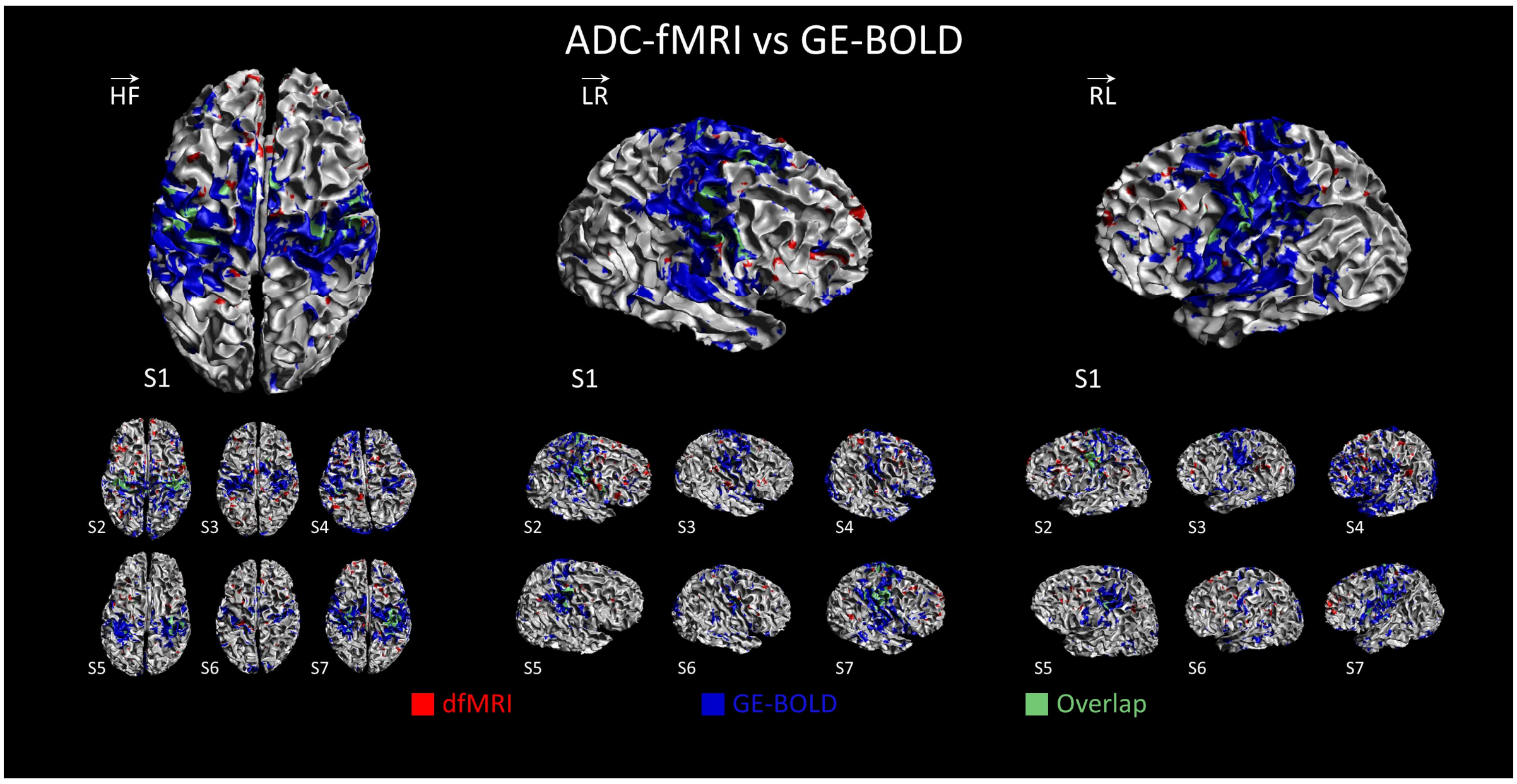

Figure 5: ADC-fMRI vs GE-BOLD activations maps. Individual activation maps detected with ADC-fMRI (red) as compared to GE-BOLD (blue), and their overlap (green) overlaid on the grey/white matter surface of each subject. Bilateral activation was observed on all subjects with both sequences. However, activations with ADC-fMRI were weaker than those with dfMRI, and generally did not include the supplementary motor cortex. Compared to GE-BOLD, activations from Z(ADC-fMRI) had smaller extension.

Time-series of GE-BOLD, ADC-fMRI, $MD_{IVIM}$-fMRI and $f_{IVIM}$-fMRI within AC(ADC-fMRI) are shown in Figure 6. The average GE-BOLD signal follows the task-rest paradigm, but to a smaller extent than in the dfMRI activations. Changes of Z(ADC-fMRI) are highly coherent with the functional paradigm, with an increase during task followed by a decrease at rest. The first run of subject S3 does not show such behaviour due to a signal discontinuity artefact approximately at the half of the acquisition. During task, significantly higher ADC values than at rest are observed, +2.17 ± 1.15% for ADC-fMRI$^{300}$ ($p = 10^{-3}$), +1.14 ± 0.57% for ADC-fMRI$^{800}$ ($p = 10^{-3}$), and +1.18 ± 0.76% ($p = 10^{-3}$) for ADC-fMRI$^{1200}$. The average changes of the considered metrics from task to rest within AC(ADC-fMRI), and their statistical significance, are reported in Table 2. SE-BOLD showed significant increases in AC(ADC-fMRI) during task execution, whereas dfMRI at b = 1200s/mm$^2$ significantly decreased in the same condition.

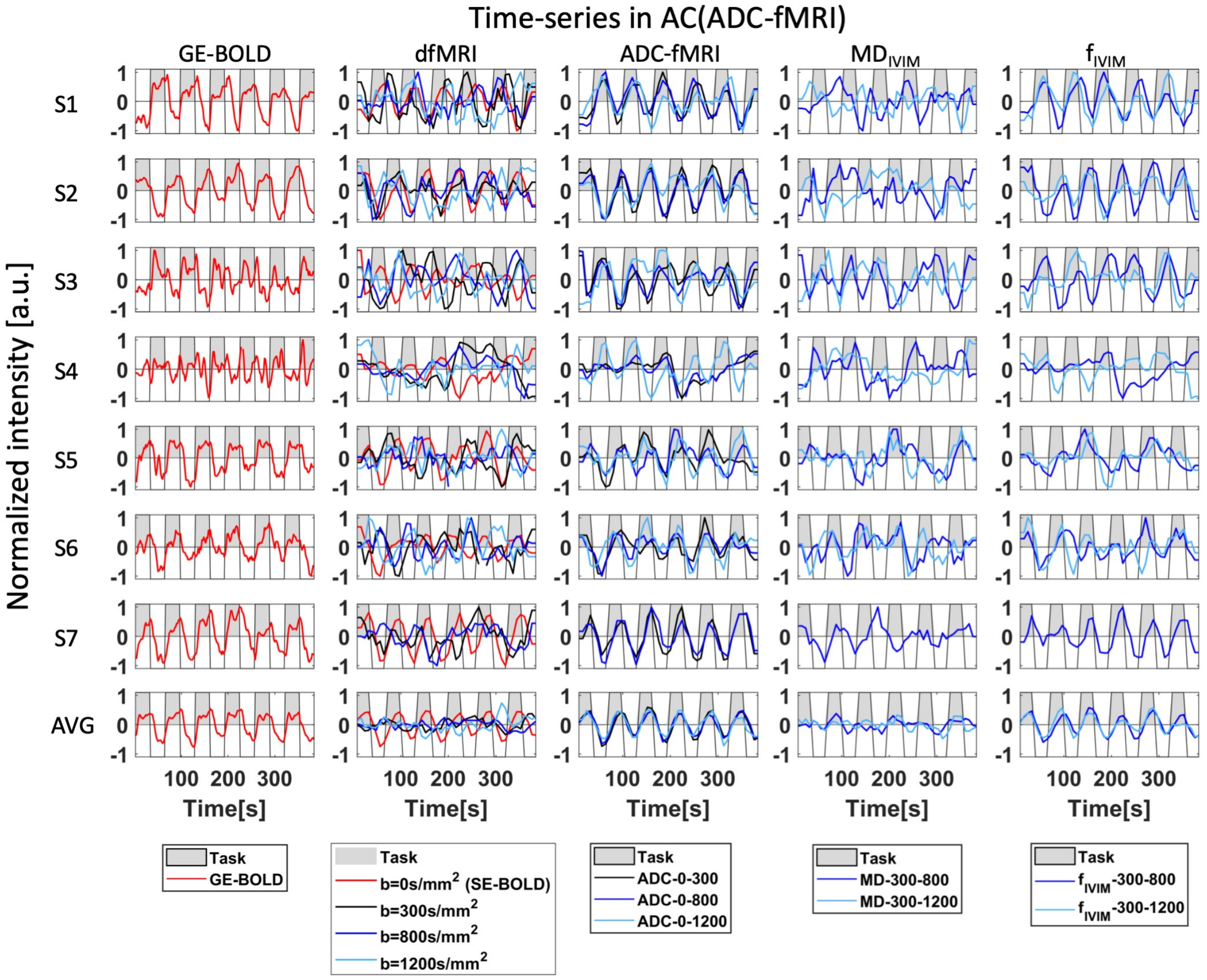

Figure 6: Time-series in AC(ADC-fMRI). Normalized average time-series of GE-BOLD, dfMRI (for different diffusion weightings, ADC-fMRI, $MD_{IVIM}$ and $f_{IVIM}$ of each subject (each row), and after group averaging (last row), within the ADC-fMRI activation ROIs. Grey blocks correspond to task execution, whereas white blocks to rest. GE-BOLD and ADC-fMRI showed increases and decreases in correspondence of task and rest, respectively. $f_{IVIM}$ increases with task execution and decreases during rest were more prominent within ADC-fMRI activations than within AC(dfMRI). The average of the $MD_{IVIM}$ signal did not exhibit clear correlations to the task.

Within the AC(ADC-fMRI) ROIs, $f_{IVIM}$-fMRI$^{300\text{-}800}$ shows a strong dependence on task execution and shows an increase of on average around 7-9% during activation compared to rest. However, the overlap between ADC-fMRI activations and $f_{IVIM}$-fMRI$^{300\text{-}800}$ activations is relatively modest, 19±11%. No correlation between $MD_{IVIM}$-fMRI and task execution is observed within AC(ADC-fMRI).

Figure 7 shows the Z-score of the time-series of GE-BOLD and ADC-fMRI computed with three different combinations of diffusion weightings. The ADC-fMRI series showed signals increases and decreases up to ± 2 standard deviations, but their changes exhibited a consistently higher delay with task execution compared to GE-BOLD and dfMRI (Figure 4).

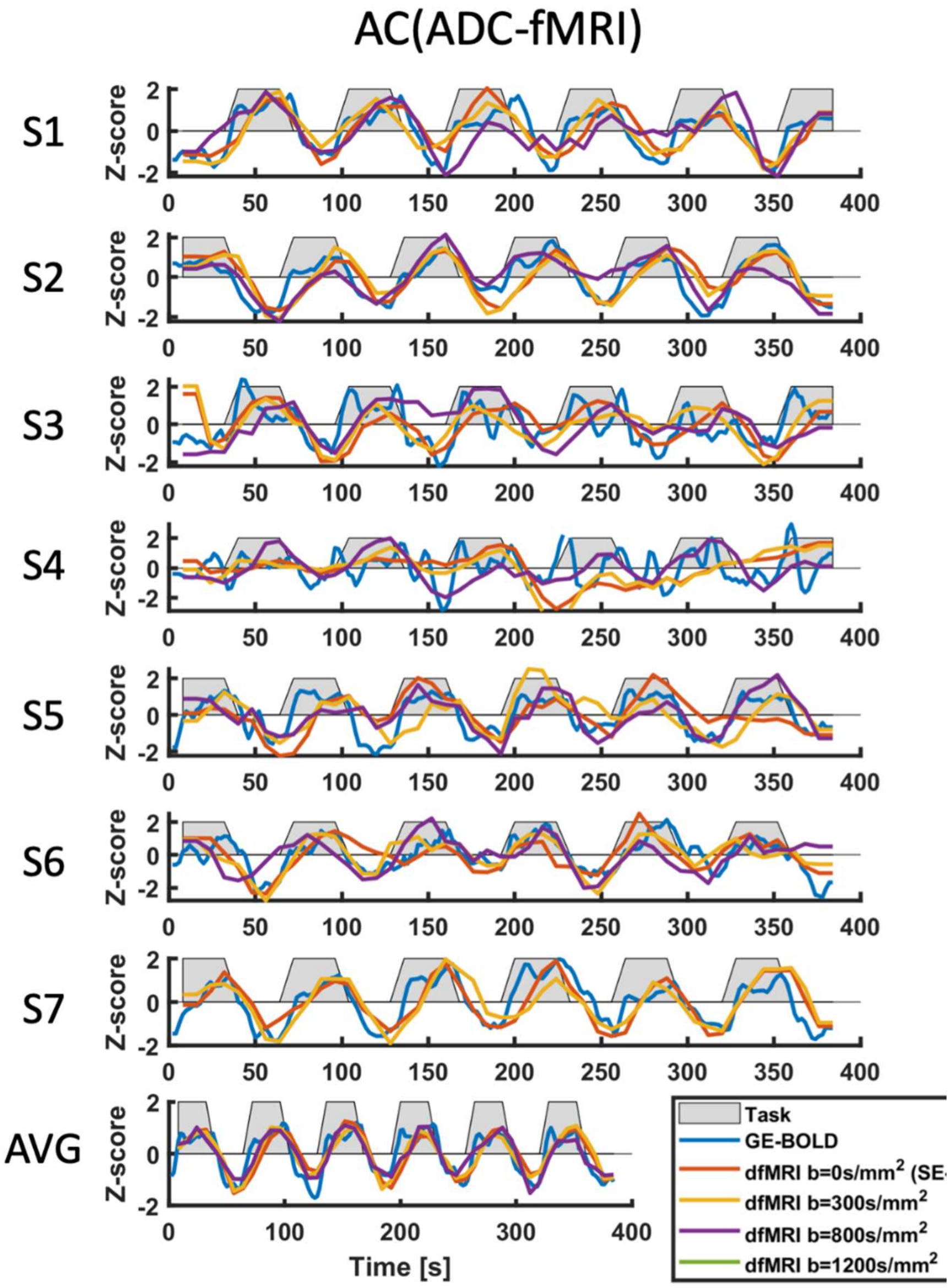

Figure 7: Sensitivity of GE-BOLD and ADC-fMRI in AC(ADC-fMRI). Z-scores of the time-series of GE-BOLD and ADC-fMRI within the ADC-fMRI activation core. The individual ADC-fMRI signals showed signal changes up to 2 standard deviations in correspondence of task execution in line with GE-BOLD, irrespectively of the applied diffusion-weighting. The time-series suggest that ADC-fMRI signals have higher delay to task execution than GE-BOLD.

## Discussion

In this work, we investigated changes in the diffusion-weighted signal measured in the brain during a motor cortex paradigm to verify whether i) the dMRI signal acquired at different diffusion weightings is sensitive to changes evoked by execution of a motor task; and ii) it is possible to separate the signal sources from the observed dfMRI signal changes. Our findings suggest that the dMRI signal is sensitive to task evoked changes in brain function, and that perfusion/free-water changes can explain only part of the observed signal alterations.

In all subjects we detected clusters where the paradigm significantly explained the dfMRI signal. The activations detected with dfMRI (Figure 2) are located approximately in the same areas as observed with GE-BOLD, but their extension is markedly smaller, in line with a previous study on the visual cortex [Nicolas et al., 2017] and with previous reports comparing Spin-Echo based activation with GE-BOLD [Glielmi et al., 2010; Norris et al., 2002]. Furthermore, the overlap between the activations from the two techniques is limited, with values smaller than 50%, suggesting adjacent and only partially overlapping areas. The reduced extension can be explained either by the reduced temporal resolution and the lower number of samples of dfMRI as compared to GE-BOLD, or by the choice of the box-car response function in place of a more sophisticated formulation [Aso et al., 2009]. Additionally, the signal crushing effect of the diffusion gradients on the vascular network may also reduce the activation extent and thus to lead to a more accurate spatial localization, as previously suggested [Song et al., 2002]. However, such effect is unlikely to be due to SNR, as dfMRI and GE-BOLD were characterized by similar SNR and tSNR levels, both around 30.

The time-series reported in Figure 3 show remarkable synchronization between task execution and dfMRI signal changes at different diffusion weightings. Although the observed alterations are modest in absolute value, i.e., below 1%, they are consistent across subjects, and statistically significant ($10^{-4}<p<10^{-2}$). Such changes are of the same order of magnitude and sign of those originally reported by Le Bihan and colleagues [Le Bihan et al., 2006], and suggest that dfMRI at weak diffusion weighting (b = 300 s/mm$^2$) is less sensitive to activation than both SE-BOLD (b = 0 s/mm$^2$) and dfMRI at intermediate to strong weighting (b = 800, 1200 s/mm$^2$). In our study, the average signal changes observed at b = 800 and 1200 s/mm$^2$ slightly anticipated changes in SE-BOLD, but to a small extent, while previous works [Aso et al., 2009; Aso et al., 2013; Le Bihan et al., 2006] reported that the time-to-peak of the dfMRI signal is considerably faster than both SE-BOLD and GE-BOLD. This might be explained by taking into consideration the repetition time employed in this work, 8s, which is the same order of magnitude of the time offset between dfMRI and SE-BOLD reported in the above-mentioned works. It should also be noted that in contrast to the abovementioned studies, in this study we acquired two diffusion weightings (in addition to the b = 0 s/mm$^2$ volume) per run. This results in an intrinsic temporal offset of 1.5 seconds between different diffusion weightings, which we did not take into account. While considering such shift is not trivial, this acquisition scheme offers the advantage of acquiring multiple diffusion weightings in a reasonable time and ensures that all data experience the same task evoked condition.

A controversial point about dfMRI relates to its underlying physiological mechanisms. Some studies suggest that these mechanisms have a neuronal (or closer to) origin [Le Bihan et al., 2006; Darquié et al., 2001; Nunes et al., 2019], whereas others show a perfusion related origin [Miller et al., 2007; Rudrapatna et al., 2012]. Although intermediate diffusion gradients (i.e., b = 1200 s/mm$^2$) are expected to completely suppress the perfusion signal, this might still affect the measured signal via the "$T_2$ shine-through" effect [Provenzale et al., 1999]. This would be especially true if the signal would originate from a single tissue component, which is not likely to be the case. To investigate this, we tailored our dfMRI acquisition design to disentangle multiple components. When using a two-component model, we observed increases of perfusion signal fractions in AC(dfMRI), between 3 and 4%. Hypothesizing invariance of the intra/extracellular environment, such perfusion changes would result in net signal alterations at b = 800 and 1200 s/mm$^2$ above 1%. Taking into account that $MD_{IVIM}$-fMRI (Figure 3) – which is theoretically less influenced by perfusion/$T_2$/$T_2$* contamination – shows weak but consistent decreases in AC(dfMRI) during task execution, we suggest that in such ROI a reduction of hindered diffusivity [Le Bihan et al., 2006; Darquié et al., 2001] takes place simultaneously to increases in $T_2$ and blood volume [Miller et al., 2007]. Interestingly, we observed an increase of ADC-fMRI$^{300}$ in AC(dfMRI), which seems in disagreement with previous studies. However, when considering the perfusion corrected IVIM model, a significant ADC decrease ($MD_{IVIM}$-fMRI$^{300\text{-}800}$) is revealed. Such result was previously observed also by Yacoub et al. [Yacoub et al., 2008], who reported ADC increases during activation when computing ADC from b = 1, 600 s/mm$^2$, in contrast to decreases when using data at b = 1200, 2400 s/mm$^2$. This is also consistent with the observation of a slightly faster responsiveness of the dfMRI signal to the neuronal stimulus at b = 1200 s / mm$^2$ compared to that at b = 0 s/mm$^2$ (Figure 4), and with our results on the overlap between Z(dfMRI) and Z($f_{IVIM}$-fMRI$^{300\text{-}800}$) activations, which is rather poor (around 15%), suggesting that perfusion is a potential contributor of the observed dfMRI changes. We further observed that dfMRI activations were mostly located in GM but also had a large component in WM, where swelling has been shown as a mechanism implicated in neuronal transmission [Fields, 2011]. Changes in $T_2^*$ appear to have a limited influence on the dfMRI signal, provided that the detected activations were not fully included in those derived from GE-BOLD but rather exhibited an alternative spatial pattern, as suggested by the limited overlap values equal to 45 $\pm$ 14%. This is in good agreement with the study of Tsurugizawa et al. [Tsurugizawa et al., 2013], which compared GE-BOLD and dfMRI in animals under strict experimental conditions. That study showed that the activation maps detected with GE-BOLD extend well beyond those derived with dfMRI, but also that the sensitivity of the latter to activation is preserved after the removal $T_2/T_2^*$ changes by inhibition of neurovascular coupling mechanisms.

Analysis of significantly task-activated ADC-fMRI voxels resulted in bilateral clusters partially overlapping those obtained with GE-BOLD (Figure 3), similarly to what was observed for dfMRI. Activations with ADC-fMRI are generally more confined than those from dfMRI but show less artefactual areas. Further, the supplementary motor cortex is hardly revealed, which might be partially due to the smaller sample size of ADC-fMRI series compared to dfMRI, as well as to its remarkably lower SNR values (see Table 2). Future studies performing ADC-fMRI should consider the combined penalty of reduced sample size and SNR in comparison to dfMRI, for instance, by increasing the number of collected samples. The ADC-fMRI values increase during activation between 2% (ADC-fMRI$^{300}$) and 1% (ADC-fMRI$^{800}$, ADC-fMRI$^{1200}$), which is in line with what is observed in the visual cortex by Nicolas et al. [Nicolas et al., 2017], and in the same magnitude but opposite sign of what is originally reported by Darquié et al. [Darquié et al., 2001]. Considering that $f_{IVIM}$-fMRI but not $MD_{IVIM}$-fMRI (Figure 5) show significant changes during task execution in AC(ADC-fMRI), we suggest that the observed increase in ADC values is largely driven by blood volume and $T_2$ values changes. This result might seem counterintuitive, given the dependency of ADC from $T_2$ values cancels out in the ADC equation. However, such assumption holds only if the signal originates from a single water pool, i.e., adhering to the classic mono-exponential diffusion equation. When applying the bi-exponential IVIM model, our results indeed show an increase in perfusion signal fraction between 6 and 7% in the ADC-fMRI activations (Table 2), which is more than what was observed for the dfMRI activations. The activation overlap with $f_{IVIM}$-fMRI activations is higher with ADC-fMRI (~19%) than with dfMRI (~15%) (see also Supporting Figure S1), and $f_{IVIM}$-fMRI exhibits a stronger correlation with task execution in AC(ADC-fMRI) than in AC(dfMRI). $MD_{IVIM}$-fMRI, which is theoretically less affected by perfusion effects, shows small but significant decreases during activation in AC(dfMRI), in line with previous reports on ADC during activations [Le Bihan et al., 2006; Darquié et al., 2001; Tsurugizawa et al., 2013; Yacoub et al., 2008], but not in AC(ADC-fMRI). Lastly, changes in ADC-fMRI (Figure 7) exhibit a visible lag to task execution compared to both GE-BOLD and dfMRI (Figure 4), suggesting the latter to be closer to the early activation mechanisms. Given these observations and taking into account that ADC-fMRI activations equally cover GM and WM, we conclude that perfusion is likely to be a stronger contributor in the ADC-fMRI response compared to the dfMRI response.

Some limitations of this work must be acknowledged. Our sequence design allows to repeatedly acquire data at multiple diffusion weightings in reasonable times, allowing to simultaneously derive

ADC measures and perfusion signal fractions. However, due to the inherent delay of 7s introduced between the b = 0 s/mm$^2$ volume and the last diffusion weighted volume acquired in each dynamic, the temporal resolution advantage of diffusion weighted data at strong b-value might be partially compromised if no further corrections are considered. However, the nature of our experiments does not allow to thoroughly investigate the temporal aspects of the dfMRI signal, which needs further investigation with dfMRI data acquired with shorter repetition times. The perfusion changes observed in this work are noticeably smaller than those reported in Federau et al. [Federau et al., 2015]. This might be due to simultaneous mechanisms taking place in the free water regime, which was suppressed with a fluid attenuated inversion recovery acquisition. Unfortunately, such acquisition is not advantageous in the dfMRI context due to the need for short repetition times. To further investigate such hypothesis, the dfMRI acquisition should be modified to accommodate a third intermediate diffusion weighting value within TR limitations. It is also worth mentioning that in Federau et al. (2015), the value of the TR was 12 times longer than the one used in this study, which affects the T1-weighting of the signal and, hence, may partially explain the observed differences. The acquisition of more diffusion weightings would also allow to employ more sophisticated IVIM fit approaches than the one here used, such as stretched exponentials [Koh et al., 2011] or proper multi exponential fit [van Baalen et al., 2017; De Luca et al., 2018], taking into account the diffusion coefficient of the pseudo-diffusion pool and improving fit quality.

The choice of the gradient waveform employed in a dfMRI experiment is to date not standardized and represents a further source of variability in the reported results. Previous studies have indeed employed monopolar Stejskal-Tanner gradients [Darquié et al., 2001; Nicolas et al., 2017; Yacoub et al., 2008], twice refocused spin-echo (TRSE) acquisitions [Aso et al., 2009; Aso et al., 2013; Le Bihan et al., 2006; Kohno et al., 2009; Miller et al., 2007; Tsurugizawa et al., 2013; Williams et al., 2014], as well as less conventional waveforms [Nunes et al., 2019; Song et al., 2002]. In this study, we have employed classic monopolar Stejskal-Tanner diffusion gradients, which are potentially sensitive to interactions with background gradients [Pampel et al., 2010]. The interaction between the applied diffusion gradients and background contributions is effectively removed only when using bipolar or asymmetric gradient designs [Froeling et al., 2015], but has been predicted by Pampel et al. [Pampel et al., 2010] to be attenuated also by twice refocused spin-echo (TRSE) acquisitions [Reese et al., 2003]. In particular, background gradients in presence of de-oxygenation have been shown to strongly decrease ADC values [Does et al., 1999], while simulations [Pampel et al., 2010] predict ADC values from monopolar gradients to be much less sensitive than TRSE to activation changes, configuring it as a potential confounder in our results. Nevertheless, we observed remarkable similarities between the above-mentioned studies and our findings of i) reduced activation extent and limited overlap of dfMRI as compared to GE-BOLD [Nicolas et al., 2017; Song et al., 2002] ii) dfMRI signal increase and $MD_{IVIM}$ decrease during activation [Le Bihan et al., 2006].

The results presented here are in general agreement with recent literature, suggesting potential advantages of dfMRI over GE-BOLD, especially in terms of spatial localization of the brain activated areas. However, it remains unclear whether the more focal activations detected with dfMRI (and ADC-fMRI) are closer to the real neuronal source of motor activation than conventional GE-BOLD fMRI activations, as recently suggested [Nunes et al., 2019].

Furthermore, we observed that both dfMRI and ADC-fMRI provide bilateral activations when comparing task and rest conditions, but that the overlap among the two is rather low, suggesting the detection of adjacent but not identical clusters. Establishing which of the two is closer to the neuronal activation will require further investigation, although our results suggest dfMRI as the most likely.

In conclusion, this study shows that the dfMRI signal is sensitive to task evoked activity, and that by employing appropriate diffusion weightings it is possible to investigate changes in different tissue domains. While dfMRI (and ADC-fMRI/IVIM-fMRI) is more challenging to perform and prone to artefacts than GE-BOLD, its selective sensitivity to different microstructural features has the potential to provide additional insights into brain activation mechanisms, to complement standard GE-BOLD.

## Acknowledgements

The research of A.L. is supported by VIDI Grant 639.072.411 from the Netherlands Organisation for Scientific Research (NWO). A.D.L. would like to thank NVIDIA Corporation for the donation of an NVIDIA Titan XP under the GPU grant program, which eased data processing and rendering.

# Figure captions

Figure 1: Workflow of this study. dfMRI data are firstly processed to attenuate motion and eddy currents related artefacts, then warped to the individual GE-BOLD space, which is used as standard space for all analysis. Data are geometrically averaged per diffusion weighting and used 1) directly for activation mapping or 2) to derive ADC-fMRI and IVIM-fMRI. After activations are individually mapped, temporal series of the signals in the activation ROIs are computed. Acronyms: $g_{x,y,z}$: diffusion gradient along the x, y or z axis; FWE: family-wise error; $TTP_{1,2}$: time-to-peak of the two Gamma functions.

Figure 2: dfMRI vs GE-BOLD activation maps. Individual activation maps detected with dfMRI (red) as compared to GE-BOLD (blue), and their overlap (green) overlaid on the grey/white matter surface of each subject. Bilateral activation as response to finger tapping was observed on all subjects with both sequences. Activations with dfMRI were smaller than those with GE-BOLD. Spurious activations due to multi-band reconstruction artefacts can be spotted on the dfMRI activations of some subjects.

Figure 3: Time-series in AC(dfMRI). Normalized average time-series of GE-BOLD, dfMRI (for different diffusion weightings, red, black, blue and light blue solid lines), ADC-fMRI, $MD_{IVIM}$ and $f_{IVIM}$ of each subject (each row), and after group averaging (last row), within the dfMRI activation ROIs. Grey blocks correspond to task execution, whereas white blocks to rest. GE-BOLD and dfMRI showed increases and decreases in correspondence of task and rest, respectively. $f_{IVIM}$ showed synchronization with the task execution, but to a less extent than dfMRI. The $MD_{IVIM}$ signal was characterized by strong pseudo-random fluctuations, but its average variation suggested its decrease during task execution.

Figure 4: Sensitivity of GE-BOLD and dfMRI in AC(dfMRI). Z-score of the time-series of GE-BOLD and dfMRI (for different diffusion weightings, red, orange, purple solid lines) within the dfMRI activation core. The individual dfMRI signals showed signal changes up to 2 standard deviations in correspondence of task execution in line with GE-BOLD, irrespectively of the applied diffusion-weighting. The time-series suggest that dfMRI signals at b = 1200s/mm$^2$ exhibited slightly faster reactivity to task than SE-BOLD (b = 0 s/mm$^2$), whereas differences with GE-BOLD were minimal.

Figure 5: ADC-fMRI vs GE-BOLD activations maps. Individual activation maps detected with ADC-fMRI (red) as compared to GE-BOLD (blue), and their overlap (green) overlaid on the grey/white matter surface of each subject. Bilateral activation was observed on all subjects with both sequences. However, activations with ADC-fMRI were weaker than those with dfMRI, and generally did not include the supplementary motor cortex. Compared to GE-BOLD, activations from Z(ADC-fMRI) had smaller extension.

Figure 6: Time-series in AC(ADC-fMRI). Normalized average time-series of GE-BOLD, dfMRI (for different diffusion weightings, ADC-fMRI, $MD_{IVIM}$ and $f_{IVIM}$ of each subject (each row), and after group averaging (last row), within the ADC-fMRI activation ROIs. Grey blocks correspond to task execution, whereas white blocks to rest. GE-BOLD and ADC-fMRI showed increases and decreases in correspondence of task and rest, respectively. $f_{IVIM}$ increases with task execution and decreases during rest were more prominent within ADC-fMRI activations than within AC(dfMRI). The average of the $MD_{IVIM}$ signal did not exhibit clear correlations to the task.

Figure 7: Sensitivity of GE-BOLD and ADC-fMRI in AC(ADC-fMRI). Z-scores of the time-series of GE-BOLD and ADC-fMRI within the ADC-fMRI activation core. The individual ADC-fMRI signals showed signal changes up to 2 standard deviations in correspondence of task execution in line with GE-BOLD, irrespectively of the applied diffusion-weighting. The time-series suggest that ADC-fMRI signals have higher delay to task execution than GE-BOLD.

## Table captions

Table 1 – Imaging parameters of the sequences employed in this study. No slice gap was employed for any of the sequences. All functional acquisitions were performed with echo planar imaging readout. Data at b = 300, 800, 1200 s/mm$^2$ were acquired with gradients along three orthogonal directions aligned with the scanner axes. MB: multi-band; SENSE: sensitivity encoding parallel imaging acceleration; TE: echo time; TR: repetition time; FOV: field of view; PE: phase encoding.

Table 2 – Values at rest of the time-series of dfMRI, ADC-fMRI, $MD_{IVIM}$-fMRI $f_{IVIM}$-fMRI in AC(dfMRI) and AC(ADC-fMRI), and their signal change during task as compared to rest. The p-value refers to a two-sided t-test between the average values in the two conditions and is highlighted in bold when significant.

# Supporting Information

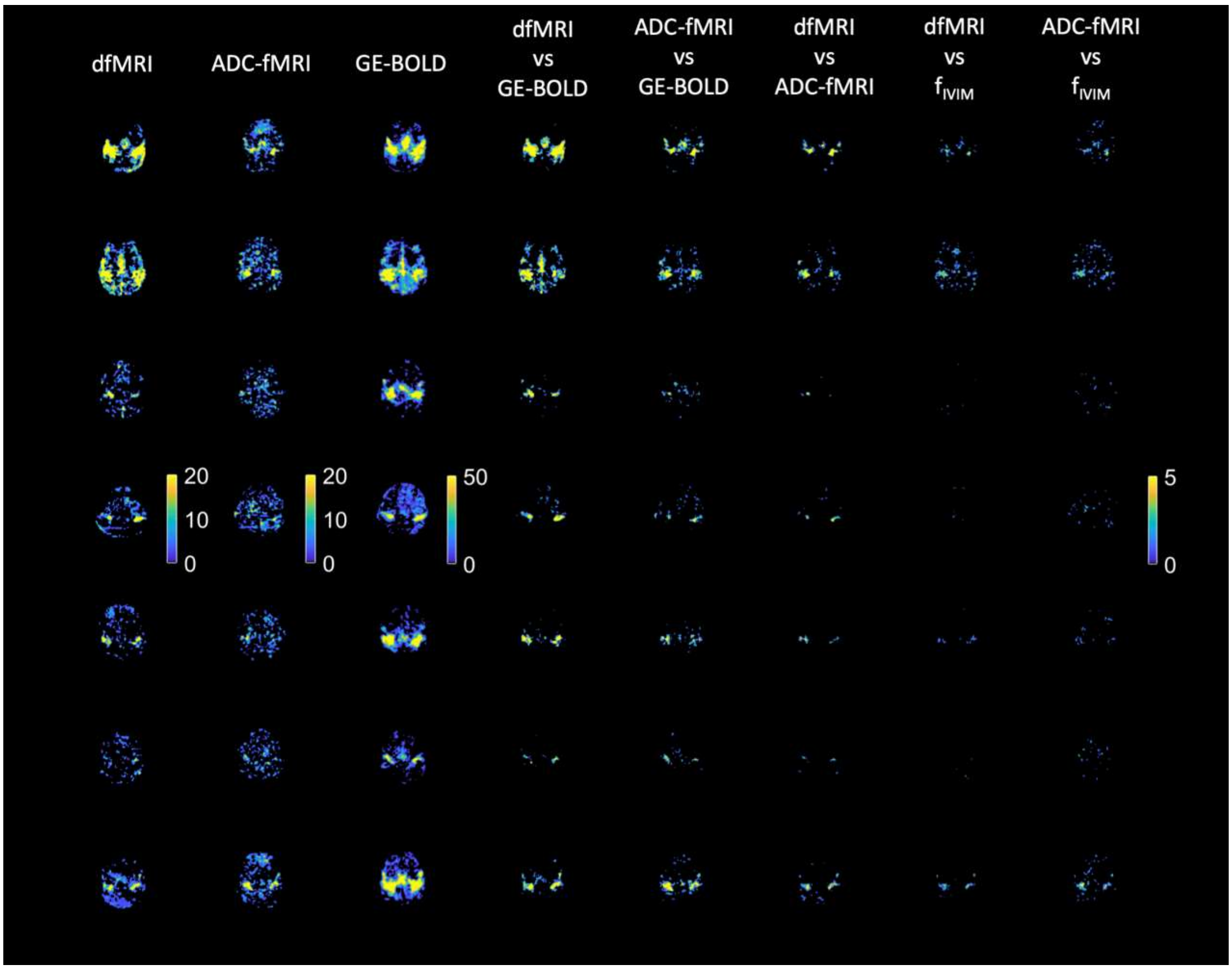

Figure S1: Axial projection of the significant Z-scores derived with dfMRI, ADC-fMRI, GE-BOLD (columns 1-3), and of their overlaps (columns 4-8). GE-BOLD resulted in the largest activation areas, followed by dfMRI and ADC-fMRI. Activations computed with ADC-fMRI visually had slightly larger overlap with $f_{IVIM}$ than those computed with ADC-fMRI.